# Quenching rates of $N_2(C^3\Pi_u, v = 0 - 4)$ states by O₂, N₂, H₂, CO₂ and CH₄

Jean-Loup du Garreau[*]

*Université Paris-Saclay, CNRS, CentraleSupélec, Laboratoire EM2C, Gif-sur-Yvette, 91190, France*

Jean-Baptiste Perrin-Terrin [†]

*Université Paris-Saclay, CNRS, CentraleSupélec, Laboratoire EM2C, Gif-sur-Yvette, 91190, France*

Christophe O. Laux [‡]

*Université Paris-Saclay, CNRS, CentraleSupélec, Laboratoire EM2C, Gif-sur-Yvette, 91190, France*

**This work presents the development of a protocol for the measurement of collisional deexcitation rate constants of $N_2(C^3\Pi_u, v = 0 - 4)$ by O₂, N₂, H₂, CO₂ and CH₄. Excited nitrogen is produced using nanosecond repetitively pulsed discharges. The applied voltage and the repetition frequencies are adjusted to minimize rotational and gas heating while keeping high electronic and vibrational excitation levels. The populations of the vibrational levels are measured using optical emission spectroscopy of the $N_2(C \to B, \Delta v = 0)$ vibrational bands. Rate coefficients for each vibrational level are then determined from the measured lifetimes. The results are compared to the available literature and new values are provided for the missing vibrational levels.**

## Nomenclature

| | | |
|---|---|---|
| $n_M$ | = | Population number density of the species M |
| $v, w, \gamma$ | = | Vibrational level |
| $I_{v'v''}$ | = | Intensity of the $N_2(C, v') \to N_2(B, v'')$ band |
| $A_{v'v''}$ | = | Einstein coefficient of the radiative transition $N_2(C, v') \to N_2(B, v'')$ |
| $\Delta E_{v'v''}$ | = | Energy difference between $N_2(C, v')$ and $N_2(B, v'')$ |
| $k_{q,M}^v$ | = | Quenching rate constant of $N_2(C, v')$ by M |
| $V_{peak}$ | = | Peak voltage of the discharge |
| $\tau$ | = | Lifetime of excited populations |
| $T_{rot}$ | = | Rotational temperature of a molecule |
| $T_{vib}$ | = | Vibrational temperature of a molecule |
| $C$ | = | Spectral + geometrical light collection attenuation of the system |

## I. Introduction

In non-equilibrium discharges in air, the main electron energy loss processes correspond to the excitation of excited electronic states of N₂. The chemical effect of the plasma on the gas is driven mostly by the collisional deexcitation or quenching of these excited states by the gas particles. The rate constants of these reactions are well known for lower vibrational levels in pure nitrogen or in air. However, in many applications, nitrogen or air is mixed with other gases. One such application is plasma-assisted combustion (PAC) in which N₂ can be mixed with O₂ and the fuel in the fresh gases and with unburnt fuel, CO, $CO_2$ and $H_2O$ in the burnt gases. Additionally, the decarbonation of combustion involves new fuels such as $H_2$ or $NH_3$ and the quenching rate constants with these molecules are often not well known. These rate constants are important for modeling the effect of plasma on the system, both in simplified [1–3] and detailed models [4–6]. The rate constants are not always available in the literature or vary greatly from one source to another, as shown in Table 1. In certain cases, the various literature sources differ by more than a factor 2: in the case of $k_{q,N_2}^0$, for instance, Bak et al. [7] found $2.5 \times 10^{-11}$ $cm^3.s^{-1}$ vs. $1.14 \times 10^{-11}$ $cm^3.s^{-1}$ for Dilecce et al. [8]. New measurements are therefore necessary. In this work, we measure these rate constants using Nanosecond Repetitively Pulsed (NRP) discharges as a plasma source.

---

[*] Ph.D. student.
[†] Ph.D. student.
[‡] Professor, AIAA Fellow.

Table 1  State-of-the-art of the quenching rate constants of $N_2$(C, v = 0-4) by $N_2$, $O_2$, $CO_2$, $H_2$ and $CH_4$, given in $10^{-10}$ $cm^3s^{-1}$.

| Quencher \ v level | v = 0 | v = 1 | v = 2 | v = 3 | v = 4 | References |
|---|---|---|---|---|---|---|
| $N_2$ | 0.071 – 0.32 | 0.096 – 0.33 | 0.26 – 0.63 | 0.428 - 0.8 | 0.49 – 0.88 | [7–25] |
| $O_2$ | 1 – 3.12 | 3.1 – 3.11 | 3.7 | 4.3 | - | [7,9,11,13,14,18,20,24] |
| $CO_2$ | 3.6 | 3.4 | - | - | - | [14] |
| $H_2$ | 3.2 – 3.3 | 3.2 | 3.7 | 4.3 | - | [18,20] |
| $CH_4$ | 5.08 – 6.53 | 6.0 | - | - | - | [14,25] |

## II. Experimental setup

NRP discharges are applied between two electrodes placed in a vacuum cell called "quenching reactor" as shown in Fig 1. The voltage and current traces are measured with electric probes (Lecroy PPE 20kV and Pearson6585) connected to an oscilloscope (TeledyneLecroy HDO6104A) as shown in Fig. 2. The gas pressure in the quenching reactor is measured by a capacitance manometer (MKS Baratron 626) and controlled by a pump (VacuuBrand MV 2 NT) and a valve. When the valve is closed and the pump is on, the pressure drops to a few millibars. The composition of the gas can be controlled with two mass flow controllers (Bronkhorst F-201AC-AAA-33-V for $N_2$ and Bronkhorst F-201CV-10K-RAD-33-K for the quenching species). The light emitted by the plasma is focused by two parabolic mirrors onto the entrance slit of a spectrometer (Acton Spectra 500i). The spectrometer is equipped with three gratings (600 gr/mm, 1200 gr/mm, and 2400 gr/mm) placed on a rotating turret. A gated ICCD camera (Princeton Instruments PI-MAX 4) is mounted on the exit of the spectrometer to measure emission spectra with 2ns gates.

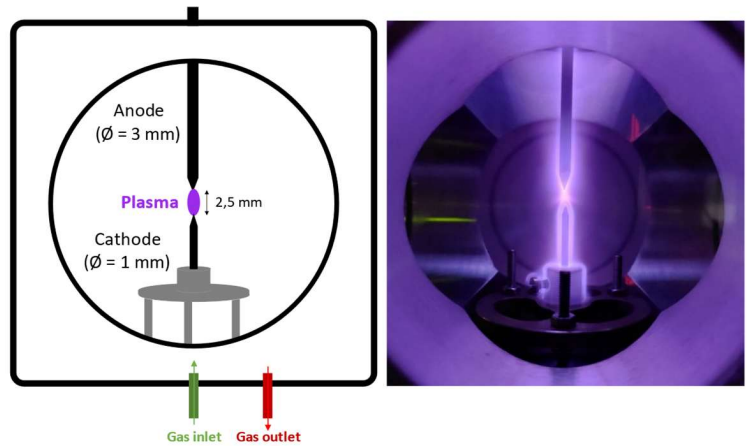

Fig. 1  Schematic of the quenching reactor (left); photograph of an $N_2$ plasma in the reactor (right).

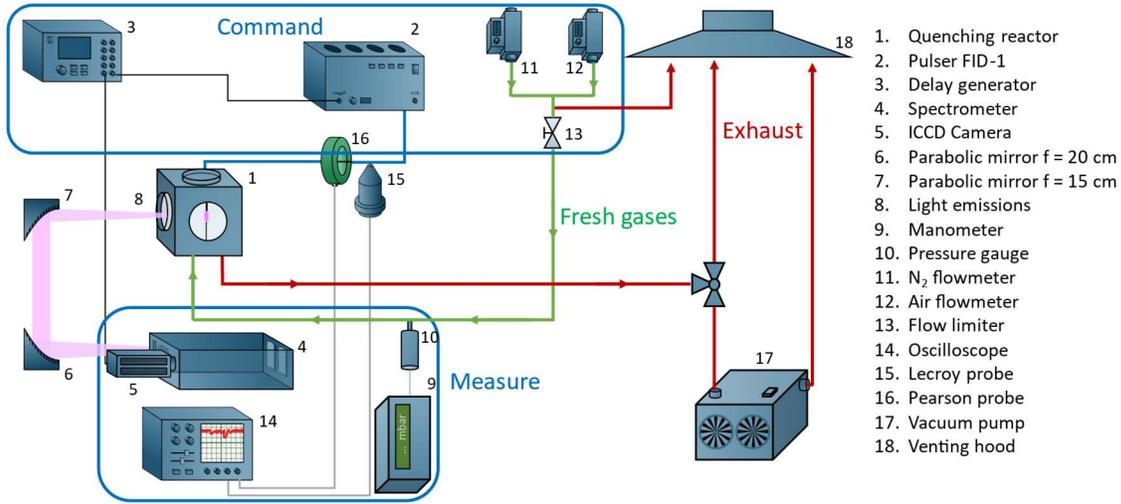

Fig. 2 Diagram of the experimental setup, N₂ + Air configuration

The high-voltage pulses used to generate plasma are delivered by an FID Technologies (model FPG 10-30MS) pulse generator with a rise time of 5 ns and a full-width at half maximum of 10 ns. The generator is externally triggered by a Pulse-and-Delay Generator BNC model 575 8C-H, enabling frequency control and time delay adjustment with the camera. In this work, the peak voltage is between 900 V and 2.5 kV, and the pulse repetition frequency (PRF) between 3 kHz and 20 kHz.

## III. Method

### A. Determination of quenching reaction rates using NRP discharges

There are two distinct phases when applying NRP discharges. First comes the 10-ns pulse, during which a high reduced electric field (typically between 100-200 Td) creates excited species via electron collisions. Then, in the afterglow following the pulse, the electric field is no longer applied and the excited species deexcite through both radiation and collisions. In this afterglow phase when the gas no longer experiences electronic excitation, a simple model of the kinetics involved can be made.

If we consider a mixture of excited N₂ with a quencher M, representative of the state of the plasma at the beginning of the afterglow, the following reactions take place:

$$N_2(C,v) \xrightarrow{A_{vw}} N_2(B,w) + photon \quad (1)$$

$$N_2(C,v) + N_2(X) \xrightarrow{k_{q,N_2}^v} N_2^* + N_2(X,\gamma) + heat \quad (2)$$

$$N_2(C,v) + M \xrightarrow{k_{q,M}^v} N_2^* + M^* + heat \quad (3)$$

The objective is to determine $k_{q,M}^v$, assuming that the other rate constants are known. The temporal evolution of the number density of level N₂(C, v) is given by:

$$\frac{dn_{N_2(C,v)}}{dt}(t) = -n_{N_2(C,v)}(t)\left(k_{q,N_2}^v n_{N_2} + \sum_w A_{vw} + k_{q,M}^v n_M\right) \quad (4)$$

where $n_i$ is the number density of species *i*. For a small mole fraction of quencher M, it can be assumed that these reactions dominate over reactions with other species such as the subproducts of M. The number density $n_{N_2(C,v)}$ then follows an exponential decay:

$$n_{N_2(C,v)}(t) = n_{N_2(C,v)}^0 e^{-\frac{t}{\tau}} \quad (5)$$

where $\tau_v$ is the lifetime of $n_{N_2(C,v)}$, defined by:

$$\frac{1}{\tau_v} = \sum_w(A_{vw}) + k_{q,N_2}^v n_{N_2} + k_{q,M}^v n_M \qquad (6)$$

By measuring the lifetime of $n_{N_2(C,v)}$, the quenching rate constant of N$_2$(C, v) by quencher M can be determined.

## B. Determination of the lifetime

The populations of excited species are monitored by calibrated Optical Emission Spectroscopy. Spectra are calibrated in intensity using an OL455 Integrating Sphere. Fig. 3 shows typical spectra measured between 325 and 339 nm during the afterglow of a discharge in pure nitrogen at 50 mbar. The spectrum is obtained by merging two spectral ranges. The resulting spectrum representing the Δv = 0 transition series of the second positive system of N$_2$ is then fitted using Specair [26][27]. The rotational temperature, T$_{rot}$, is determined by fitting the N$_2$(C, v=0) → N$_2$(B, w=0) band, after which the bands from v=1-4 can be fitted by adjusting the populations of their respective vibrational levels to match the measurements. Fig. 3 shows the result of the fit on an experimental spectrum with a T$_{rot}$ step of 10 K. T$_{rot}$ increases from 300 K during the first 6 nanoseconds of the discharge to a stable temperature between 380 and 420 K for the remainder of the discharge and afterglow. This temperature increase depends on the pressure and discharge voltage.

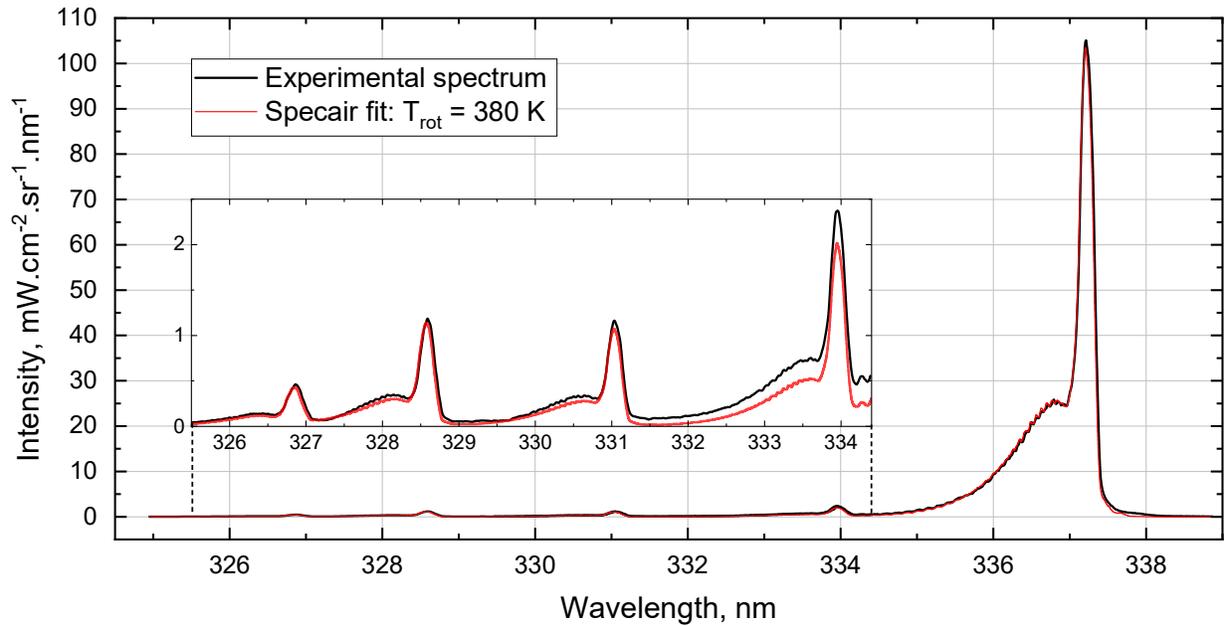

**Fig. 3 Black: Spectra measured in the afterglow of a discharge in pure nitrogen at pressure 50 mbar, with a 2400-gr/mm grating. Red: fitted spectrum of N$_2$ second positive emission calculated with Specair [26][27] at T$_{rot}$ = 380 K.**

The temporal evolution of the various vibrational level populations is obtained by fitting spectra at increasing delays from the start of the discharge, as shown in Fig. 4. The populations follow single exponential decay laws, which confirm the assumption made on $n_{N_2(C,v)}$ in the model. Computing linear regressions on the logarithm of the populations in the afterglow gives slopes corresponding to the inverse lifetimes. Quenching rate constants can be inferred from these rates as detailed below.

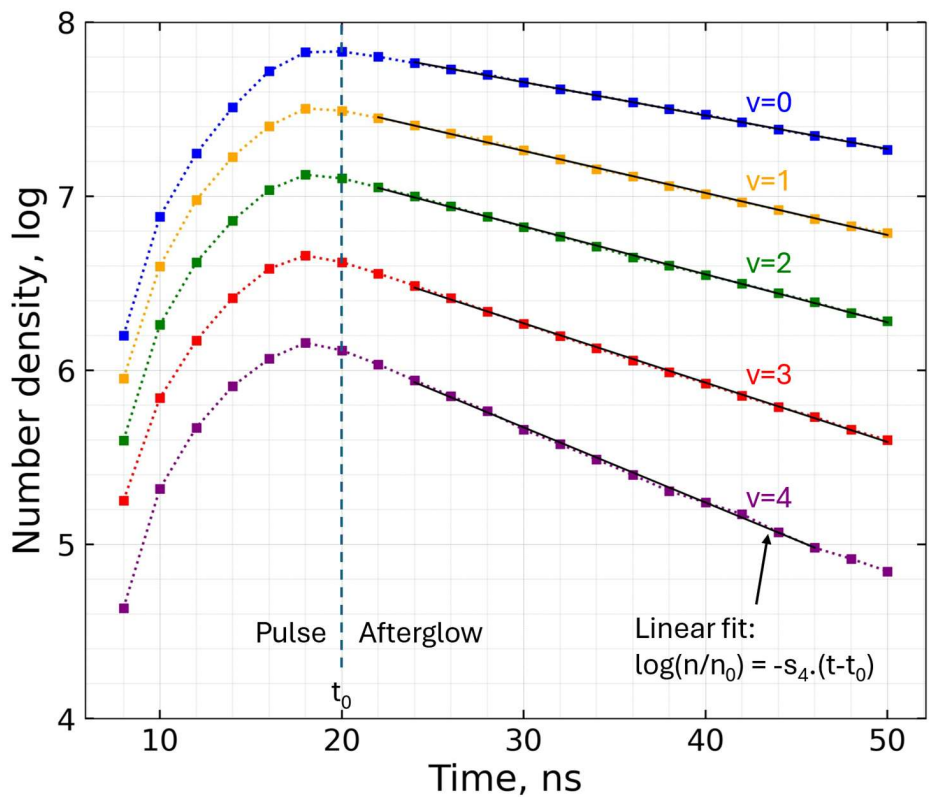

**Fig. 4  Logarithmic evolution of the vibrational populations with linear regressions.**

From these measured decay slopes $s_v$ and Eq. 6, the quenching rate constants of N$_2$(C, v) by quencher M can be inferred:

$$k_{q,M}^v = \frac{s_v - \sum_w A_{vw} - n_{N_2} k_{q,N_2}^v}{n_M} \qquad (7)$$

## C. Effect of pressure and quencher concentration

By measuring the inverse lifetimes in this way at different pressures, we obtain Stern-Volmer plots such as those shown in Fig. 5. Another method is to perform the fits for each vibrational level v while fixing the intercept at zero pressure at the value of the radiative lifetime $(\sum_w A_{vw})^{-1}$ of level v. In our case, both methods essentially give the same values.

All molecules tested are found to have quenching rate constants of N$_2$(C) that are greater (by approximately one order of magnitude) than the quenching rate constant by N$_2$. Thus, the inverse lifetimes significantly increase when the concentration of these quenching molecules is increased. An example is shown in Fig. 6, where we see that the slopes increase as the mole fraction of oxygen grows.

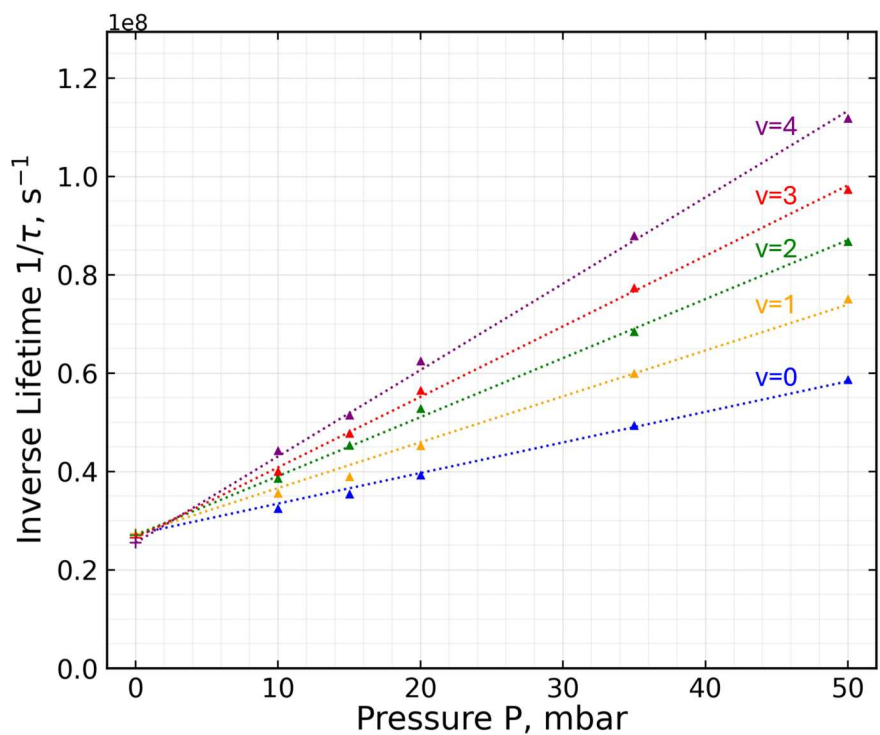

**Fig. 5 Stern-Volmer plot for the quenching of $N_2(C, v=0-4)$ in a 95:5 mixture of $N_2$ and $CO_2$.**

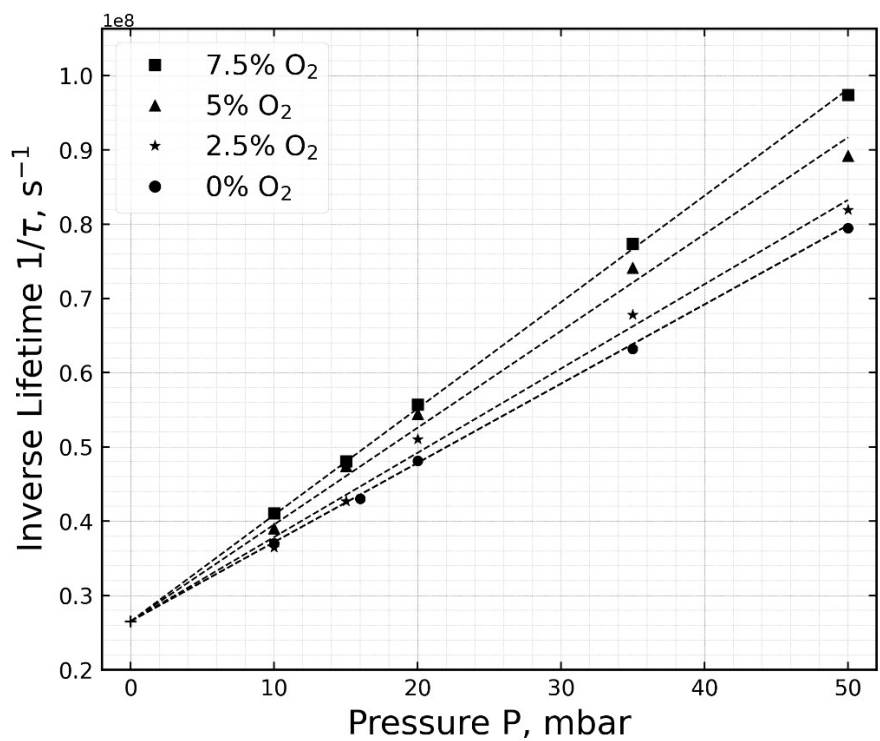

**Fig. 6 Stern-Volmer plot of $N_2(C, v=3)$ for different mole fractions of $O_2$**

## IV. Results

Measurements were made for the quenching by pure $N_2$, and with mole fractions of 2.5%, 5% and 7.5% for $O_2$, $H_2$, $CO_2$ and $CH_4$. Various concentrations led to consistent lifetimes and quenching rates. Comparative graphs of the values found by different groups [7–25] and those measured in this work are shown in Fig. 7 and Fig. 8:
- For $N_2$, as shown in Fig. 7, the current values are in good agreement with the literature, with values on the lower end of the distribution for v=3,4.
- For $O_2$, Fig. 8 shows excellent agreement for v=0-2, and good agreement for v=3. No previous measurement was found for v=4.
- For $CH_4$, Fig. 8 shows good agreement for v=0,1. No previous measurements were found for v=2-4.
- For $CO_2$, Fig. 8 shows good agreement for v=0,1. No previous measurements were found for v=2-4.
- For $H_2$, Fig. 8 shows good agreement for v=0-3. No previous measurement was found for v=4.

Table 2 gives the exact values of all the measured quenching rate constants and the measurement uncertainty.

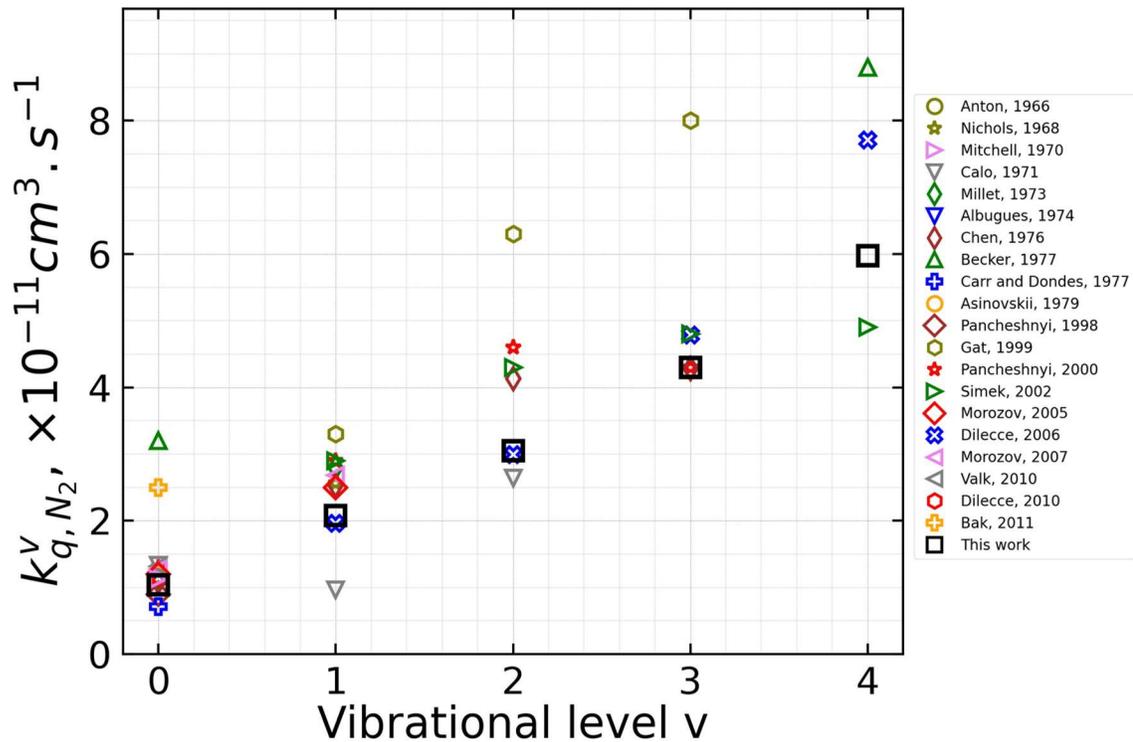

**Fig. 7 Comparison of the quenching rate constants of $N_2(C, v=0-4)$ by $N_2$.**

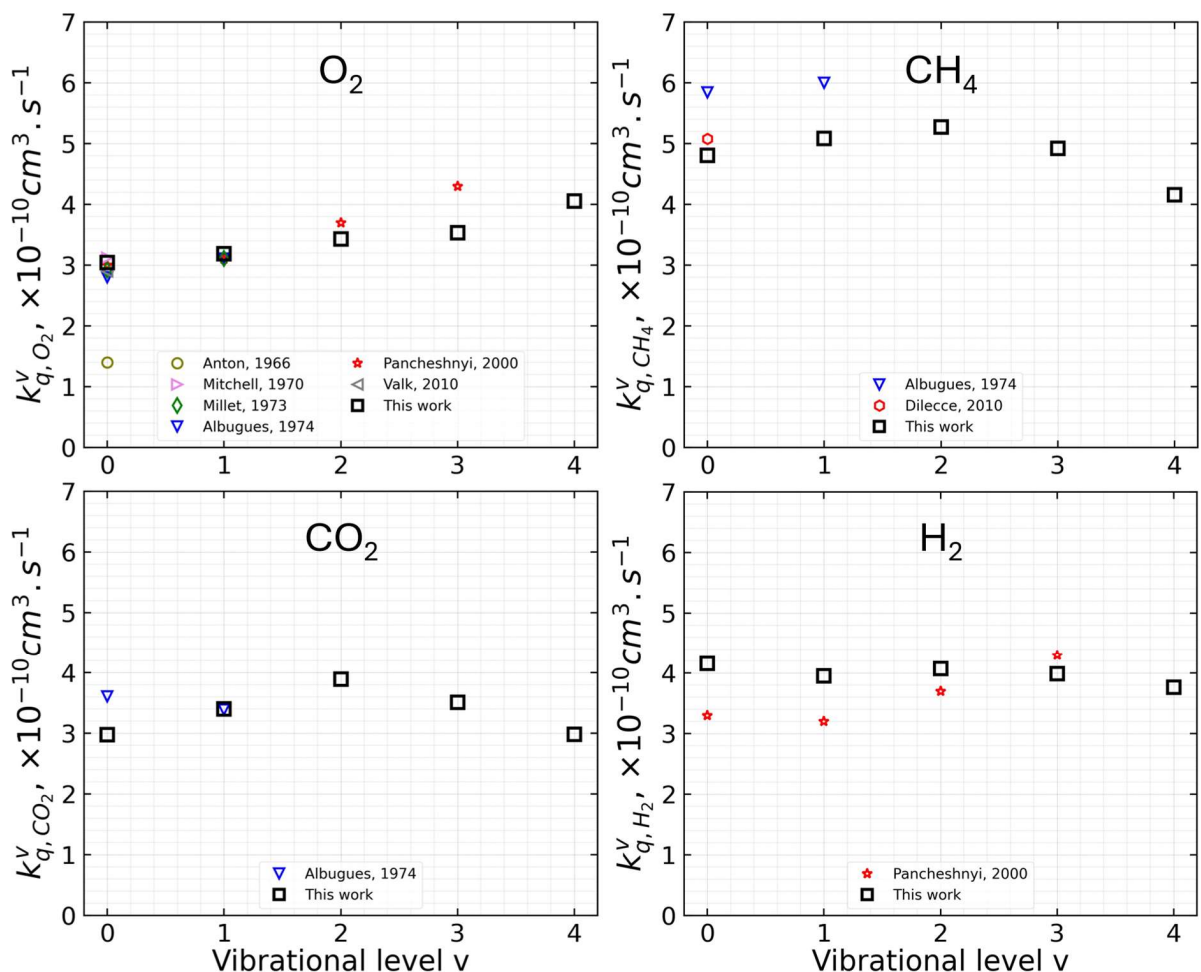

**Fig. 8** Comparison of the quenching rate constant values of $N_2(C, v=0-4)$ by $O_2$, $CH_4$, $CO_2$ and $H_2$.

**Table 2** Measured quenching rate constants of $N_2(C, v = 0-4)$ by $N_2$, $O_2$, $CO_2$, $H_2$ and $CH_4$, given in $10^{-10}$ cm$^3$s$^{-1}$.

| Quencher \ v level | v = 0 | v = 1 | v = 2 | v = 3 | v = 4 |
|---|---|---|---|---|---|
| $N_2$ | 0.10 ± 0.01 | 0.21 ± 0.01 | 0.31 ± 0.02 | 0.43 ± 0.02 | 0.60 ± 0.02 |
| $O_2$ | 3.0 ± 0.03 | 3.2 ± 0.1 | 3.4 ± 0.2 | 3.5 ± 0.2 | 4.1 ± 0.2 |
| $CO_2$ | 3.0 ± 0.1 | 3.4 ± 0.2 | 3.9 ± 0.2 | 3.5 ± 0.1 | 3.0 ± 0.1 |
| $H_2$ | 4.2 ± 0.2 | 4.0 ± 0.0 | 4.1 ± 0.1 | 4.0 ± 0.1 | 3.8 ± 0.1 |
| $CH_4$ | 4.8 ± 0.1 | 5.1 ± 0.1 | 5.3 ± 0.2 | 4.9 ± 0.1 | 4.2 ± 0.1 |

## V. Conclusions and future work

We have developed a method to measure the quenching rate constants of $N_2(C, v)$ by various species using NRP discharges. This method is validated against the literature results. Additionally, the set of quenching rate constants of $N_2(C, v =0-4)$ by the species presented in Table 1 is now extended by our measurements in Table 2, especially for the high vibrational levels of $N_2(C)$. Good agreement is obtained with the quenching rate constants measured by Albugues *et al* [14] for $CO_2$ and $CH_4$.

In future work, we will extend our results to the quenching rate constants of $N_2(C)$ by additional species. We also plan to investigate the quenching rate constants of $N_2(B)$, which plays an important role in the kinetics of non-equilibrium nitrogen-containing plasmas. We also intend to determine the products formed by collisional quenching with the relevant gases.

## Acknowledgments

This work has received funding from the European Research Council (ERC) under the European Union's Horizon 2020 research and innovation program (grant agreement No. 101021538). A CC-BY4.0 public copyright license has been applied by the authors to the present document and will be applied to all subsequent versions up to the Author Accepted Manuscript arising from this submission, in accordance with the grant's open access conditions. The authors thank Erika Jean-Bart, Yannick Le Teno and Moïses Ramires Garcia for their help in the design and construction of the setup.